\documentclass[floatfix,prd,twocolumn,letterpaper,lengthcheck,superscriptaddress,showpacs,amssymb,amsmath,amsfonts,aps,altaffilletter,longbibliography]{revtex4-1}

\usepackage{color}
\usepackage[hidelinks]{hyperref}
\usepackage{amsmath}
\usepackage{amsthm}
\usepackage{multirow}
\usepackage{graphicx}
\usepackage{tikz}
\usetikzlibrary{calc}
\usepackage{placeins}
\usepackage{xspace}
\usepackage{physics,enumerate}

\theoremstyle{conjecture}

\usepackage{mathrsfs}
\DeclareSymbolFontAlphabet{\mathrsfs}{rsfs}

\newcommand{\be}{\begin{equation}}
\newcommand{\ee}{\end{equation}}
\newcommand{\ba}{\begin{eqnarray}}
\newcommand{\ea}{\end{eqnarray}}
\theoremstyle{definition}
\newtheorem*{conjecture1*}{Conjecture 1}
\newtheorem*{conjecture2-1*}{Conjecture 2-1}
\newtheorem*{conjecture2-2*}{Conjecture 2-2}

\let\oldbibitem\bibitem
\renewcommand{\bibitem}{%
  \renewcommand{\doi}[1]{doi: ##1}
  \let\bibitem\oldbibitem
  \oldbibitem
}

\begin{document}

\title[]{Entropy Bounds for Rotating AdS Black Holes}

\author{Masaya Amo}
\affiliation{Center for Gravitational Physics and Quantum Information,
Yukawa Institute for Theoretical Physics,
Kyoto University, Kyoto 606-8502, Japan}
\affiliation{Departament de F{\'\i}sica Qu\`antica i Astrof\'{\i}sica, Institut de
Ci\`encies del Cosmos,
 Universitat de
Barcelona, Mart\'{\i} i Franqu\`es 1, E-08028 Barcelona, Spain
}
\author{Antonia M.~Frassino}
\affiliation{Departament de F{\'\i}sica Qu\`antica i Astrof\'{\i}sica, Institut de
Ci\`encies del Cosmos,
 Universitat de
Barcelona, Mart\'{\i} i Franqu\`es 1, E-08028 Barcelona, Spain
}
\author{Robie A.~Hennigar}
\affiliation{Departament de F{\'\i}sica Qu\`antica i Astrof\'{\i}sica, Institut de
Ci\`encies del Cosmos,
 Universitat de
Barcelona, Mart\'{\i} i Franqu\`es 1, E-08028 Barcelona, Spain
}

\begin{abstract}
We propose novel thermodynamic inequalities that apply to stationary asymptotically Anti-de Sitter (AdS) black holes. These inequalities incorporate the thermodynamic volume and refine the reverse isoperimetric inequality. To assess the validity of our conjectures, we apply them to a wide range of analytical black hole solutions, observing compelling evidence in their favour. Intriguingly, our findings indicate that these inequalities may also apply for black holes of non-spherical horizon topology, as we show their validity as well for thin black rings in AdS. 
\end{abstract}

\maketitle

\section{Introduction}

The thermal nature of black holes underpins many of the deepest insights into quantum gravity. Black hole entropy ensures the consistency of the second law of thermodynamics in our universe and is a `smoking gun' for a microscopic description of the gravitational field~\cite{Bekenstein:1973ur, Hawking:1975vcx}. Over the last decade, our understanding of the laws of black hole mechanics has expanded to include pressure and volume~\cite{Kastor:2009wy,Cvetic:2010jb}. The study of these terms is often called \textit{extended black hole thermodynamics} and has led to new perspectives on gravitational phase transitions~\cite{Kubiznak:2012wp}, black hole heat engines~\cite{Johnson:2014yja}, and holography~\cite{Karch:2015rpa,AlBalushi:2020rqe,Visser:2021eqk,Frassino:2022zaz, Ahmed:2023snm}.

Extended black hole thermodynamics centers around thermodynamic volume, which has a geometric definition in terms of Komar integrals and is important for rendering consistent the Smarr formula for AdS black holes~\cite{Kastor:2009wy}. If one allows for variations in the cosmological constant, then the thermodynamic volume appears in the first law as its conjugate quantity.

A particularly interesting early result in extended thermodynamics is the \textit{reverse isoperimetric inequality} (RII)~\cite{Cvetic:2010jb}. The RII conjectures that for a black hole in $D$ dimensions with horizon area $A$ and thermodynamic volume $V$ the ratio
\be\label{RII} 
\mathcal{R} \equiv \left(\frac{V}{\mathcal{V}_0} \right)^{1/(D-1)} \left(\frac{\mathcal{A}_0}{A} \right)^{1/(D-2)}
\ee
satisfies $\mathcal{R} \ge 1$.\footnote{Here $\mathcal{A}_0$ and $\mathcal{V}_0$ are the area and volume of  unit surface of constant $(t,r)$. For a sphere, $\mathcal{A}_0 = \Omega_{D-2} = 2 \pi^{(D-1)/2}/\Gamma\left[(D-1)/2\right]$, and $\mathcal{V}_0 = \Omega_{D-2}/(D-1)$.} Physically, the RII is the idea that, for a black hole with a fixed thermodynamic volume, there is a maximum possible entropy. The maximum entropy is achieved for the Schwarzschild-AdS black hole, which saturates the inequality. This allows for the following alternate interpretation of the RII: The entropy of a black hole of thermodynamic volume $V$ is no more than the entropy of the Schwarzschild-AdS black hole with the same volume, or 
\be 
A(V) \le A_{\rm Schw}(V) \, .
\ee

Support for the RII is robust. For instance, 
in~\cite{Cvetic:2010jb} the RII was found to hold for a wide variety of asymptotically AdS black holes in four and higher dimensions. Further corroboration was given in~\cite{Altamirano:2014tva}, where it was found to hold for the topologically non-trivial higher-dimensional AdS black rings. In~\cite{Dolan:2013ft} it was found that the conjecture applies to black hole and cosmological horizons in asymptotically de Sitter spacetimes, while~\cite{Gregory:2019dtq} showed the RII is amenable to the inclusion of conical deficits. Intriguingly, the results of~\cite{Frassino:2022zaz} point toward a ``quantum'' RII when semi-classical effects are accounted for.  Despite this progress, no general mathematical proof of the inequality or a precise statement of its necessary or sufficient conditions has been established, except for a few specific cases. For example, in~\cite{Feng:2017wvc}, researchers proved the RII for static black holes with planar horizons, assuming an empirically motivated formula for the thermodynamic volume and the null energy condition.

In addition to intrinsic interest as a well-supported geometric and thermodynamic inequality, the RII has led to several intriguing results. For instance, in~\cite{Johnson:2019mdp}, a negative heat capacity at constant volume was linked to a violation of the RII, implying thermodynamic instability. The authors of~\cite{Johnson:2019wcq} investigated the thermodynamic volume from a microscopic perspective and suggested that a violation of the RII in three space-time dimensions is related to over-counting of the field theory entropy in the Cardy formula. Additionally, in~\cite{AlBalushi:2020rqe}, the authors established a connection between thermodynamic volume and the complexity of formation in holography. They argued that the RII can be understood as a lower bound on the complexity of formation set by the entropy.

There are no known counter-examples to the RII for asymptotically AdS black hole solutions of Einstein gravity in 
dimension $D \ge 4$. Beyond this, there are only two cases where violations of the RII have been argued for, but neither case is completely compelling. The first example pertains to a class of `ultraspinning' black holes that arise from a singular limit of the Kerr-AdS metric~\cite{Klemm:2014rda, Hennigar:2014cfa, Hennigar:2015cja,Gnecchi:2013mja}.  The ultraspinning black hole is asymptotically \textit{locally} AdS and possesses an event horizon that is non-compact yet has finite area. In the original references~\cite{Hennigar:2014cfa, Hennigar:2015cja}, it was argued that these black holes highlight the role played by horizon topology in the statement of the RII.
However, whether they provide a legitimate counter-example to the RII was subsequently called into question~\cite{Hennigar:2018cnh, Appels:2019vow}. Essentially, the ultra-spinning limit results in a metric of reduced co-homogeneity which makes the thermodynamics as considered ill-defined. 

The second possible violation of the RII is associated with electrically charged BTZ black holes and is closely connected to the challenges of defining Komar charges in lower dimensions (see also~\cite{Frassino:2019fgr}), as well as the peculiarities of the charged BTZ solution~\cite{Frassino:2015oca}. The RII is only violated when an alternate (non-geometric) definition of the thermodynamic volume is employed. If one adheres to the geometric definition of the thermodynamic volume, then the RII is satisfied and no violation occurs~\cite{Frassino:2015oca}.\footnote{See \cite{Johnson:2019wcq} for a microscopic view of super-entropic solutions.} In summary, while the literature features potential violations of the RII, to date, there is no definitive counter-example to the conjecture.

Our purpose here is to present novel inequalities involving the thermodynamic volume that are generalizations of the RII. Our refined reverse isoperimetric inequalities (RRIIs) include angular momentum. They reduce to the standard RII when the angular momentum vanishes, but are otherwise strictly stronger statements. The strongest variant of the RRII is the following. 
\begin{conjecture1*}[Strong RRII]
For a stationary asymptotically AdS black hole of mass $M$, angular momenta $J_i$, and thermodynamic volume $V$,  the following inequality holds
\be\label{RRII} 
A(M, J_i, V) \le A_{\rm Kerr} (M, J_i, V) \, ,
\ee
where $A_{\rm Kerr}$ is the area of the Kerr-AdS black hole with the same parameters.
\end{conjecture1*}
Conjecture~1 is the statement that for fixed values of $(M, J_i, V)$ the Kerr-AdS black hole (if it exists) has maximum entropy.\footnote{The updated RII deviates from its original version as both mass and angular momentum, along with volume, determine the area in the Kerr-AdS case. Note the existence of Kerr-AdS black holes is not guaranteed for every combination of $M$, $J_i$, and $V$.}  Any deformation of the solution, e.g. through the incorporation of additional charges or matter fields, leads to a decrease in the black hole entropy. In the limit $J_i \to 0$, the Kerr-AdS area reduces to the Schwarzschild-AdS area and the RII~\eqref{RII} is recovered. 

The conjecture takes inspiration from the Penrose inequality and its stronger generalizations that incorporate conserved charges~\cite{Penrose:1973um}. Restricted to stationary spacetimes, the Penrose inequality provides a bound on the mass in terms of the area of the horizon, holding as an equality for slices of the asymptotically flat Schwarzschild black hole and as an inequality for other stationary, asymptotically flat black holes. Analogously, there exists a stronger form of the Penrose inequality that includes angular momentum. This version holds as an equality for slices of the Kerr solution and as an inequality for more general solutions~\cite{Mars:2009cj, Dain:2014xda} (see also \cite{Anglada:2017ryp,Jaracz:2018jrp,Anglada:2018czw,Lee:2021hft,Lee:2022fmc,Shiromizu:2022fqk}). 

The RII~\eqref{RII} holds as an equality for the Schwarzschild-AdS black hole, while it is an inequality under more general circumstances. In the same spirit as the stronger version of the Penrose inequality, we sought to find a generalization of the RII that holds as an equality for the Kerr-AdS black holes, and then investigate whether this relation holds more generally as an inequality. This led us to the RRII given in \eqref{RRII}.  Below, we will provide evidence in favour of this conjecture by examining a large number of examples. We also present conjectures weaker than~\eqref{RRII}, but stronger than~\eqref{RII}.

\section{Evidence for the strong RRII}

We will now present the evidence we have accumulated in favour of the conjecture~\eqref{RRII}. To streamline the discussion, the form of the metrics and relevant thermodynamic parameters have been presented in the Supplement~\footnote{The Supplemental Material contains a review of extended thermodynamics, along with a presentation of the metrics used in this analysis along with their relevant thermodynamic properties. It uses refs.~\cite{Kastor:2009wy, Cvetic:2010jb, Ashtekar:1984zz, Ashtekar:1999jx, Gibbons:2004ai, Hollands:2005wt, Durgut:2022xzw, Carter:1968ks, Hawking:1998kw, Gibbons:2004uw, Caldarelli:1999xj, Gunasekaran:2012dq, Anabalon:2018qfv, Gregory:2019dtq, Chong:2004na, Cvetic:2005zi, Cvetic:2010jb, Chow:2007ts, Gibbons:2004ai, Caldarelli:2008pz, Altamirano:2014tva}}.

Consider first the Kerr-Newman-AdS black hole, for which the extended thermodynamics was first studied in~\cite{Caldarelli:1999xj, Gunasekaran:2012dq}. For this case, the following identity holds among the thermodynamic parameters
\be\label{kerrads4} 
36 \pi M^2 V^2 - M^2 A^3 - 64 \pi^3 J^4 = 16 \pi^2 Q^2 J^2 A \, .
\ee
In particular, note that when the charge $Q=0$ the metric reduces to the Kerr-AdS$_4$ solution and the left-hand side vanishes identically. To check the validity of the RRII, we hold fixed $M$, $J$ and $V$ and study how $A$ changes as the charge $Q$ is varied. To satisfy the RRII conjecture requires that the right-hand side is non-negative. This is manifestly so and therefore the conjecture~\eqref{RRII} holds for the Kerr-Newman-AdS black hole.

For our next example, we examine the charged, rotating AdS C-metric. We validate the inequality using the Christodoulou-Ruffini mass formula from~\cite{Gregory:2019dtq}, specifically referencing (17) from that study, which yields
\be
\frac{4 \pi M^2}{S} 
\le \left ( \frac{3\pi M V}{2 S^2} - \frac{2C^2}{x^2} \right)^2
-4 \left (\frac{\pi J}{S}\right)^4\, ,
\ee
while the combination of (11) and (13) of~\cite{Gregory:2019dtq} gives
\be 
 \frac{3\pi M V}{2 S^2} - \frac{2C^2}{x^2} >   0 \, .
\ee
Combining these relations and replacing $S = A/4$ we obtain
\be 
36\pi M^2V^2 - M^2A^3 -64\pi^3J^4  \ge  0 \, .
\ee
Therefore, the RRII holds for this solution.

\begin{figure}
    \centering
    \includegraphics[width=0.45\textwidth]{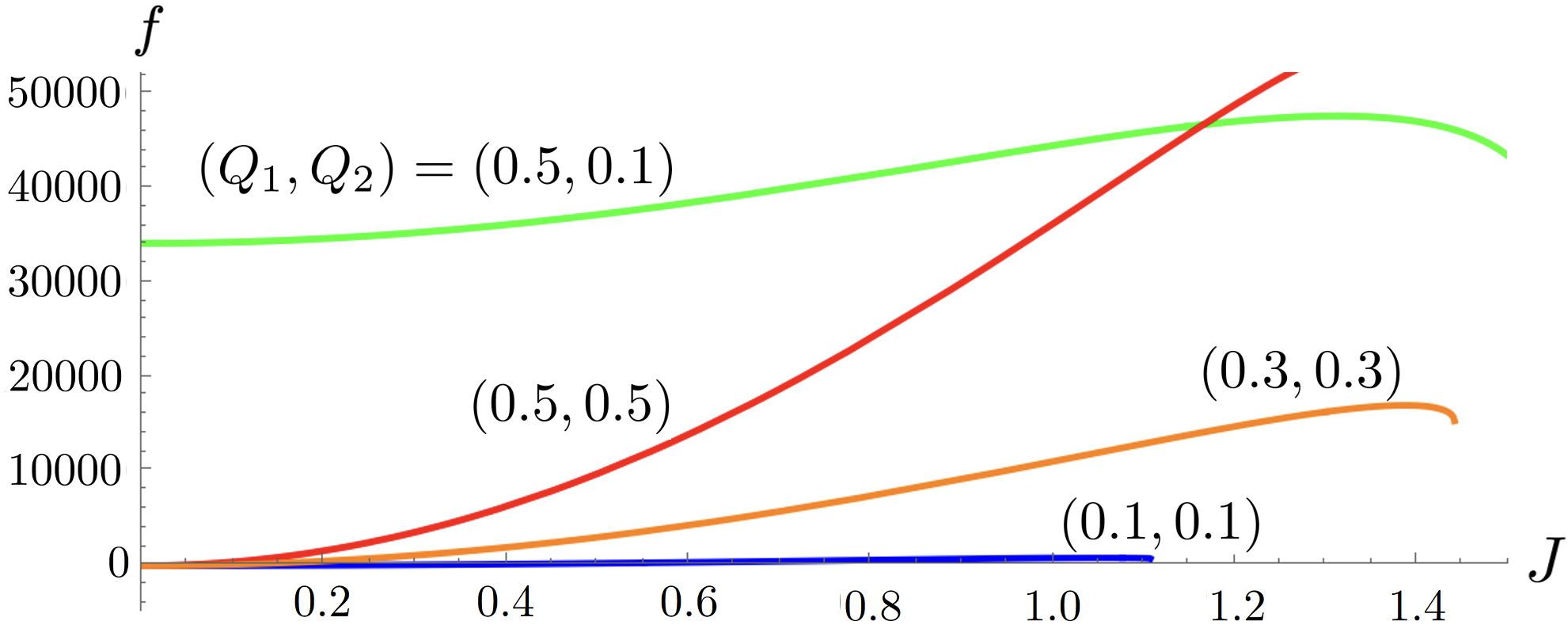}
    \caption{For the pairwise-equal charge black holes of $D=4$ gauged supergravity, we plot $f\equiv36 \pi M^2 V^2 - M^2 A^3 - 64 \pi^3 J^4$ against several charge values, keeping $m=1$ and $l=5$ constant. The RRII is valid when $f\ge0$. The curve endpoint for $(Q_1,Q_2)=(0.5,0.1)$ represents an extremal black hole.}
    \label{pairwise_equal_rrii}
\end{figure}

We next consider the rotating pairwise-equal charge black holes in $D=4$ gauged supergravity. These, characterized by two $U(1)$ charges, were first detailed in~\cite{Chong:2004na} and later analyzed thermodynamically in~\cite{Cvetic:2005zi, Cvetic:2010jb}. Using Mathematica, it is possible to directly prove that~\eqref{RRII} holds. For illustrative purposes, we present a few representative examples in Figure~\ref{pairwise_equal_rrii}.

We now turn to higher dimensions. In these cases it is not possible to proceed analytically and we resort to numerical exploration of the parameter space.   To test the RRII in five dimensions, we examine a charged and rotating solution of minimal gauged supergravity presented in~\cite{Chong:2005hr}. This solution is defined by five parameters -- $(r_+, a, b, g, q)$ -- that represent the horizon radius, spin parameters, cosmological length scale, and charge, respectively. Our testing approach involves generating random values for these parameters and verifying that they correspond to a physically reasonable black hole by ensuring a non-singular exterior metric. Next, we compute the conserved charges associated with these parameter values. Then we determine whether one (or more) Kerr-AdS$_5$ black hole exists with the same volume and conserved charges, and compute the associated parameters $(\tilde{r}_+,\tilde{a},\tilde{b},\tilde{g})$.  Finally, we compare the areas to confirm the validity of the conjecture~\eqref{RRII}.

For the solution of~\cite{Chong:2005hr}, we have carried out this procedure for approximately 100,000 randomly sampled parameter values, and in no case have we found a violation of the RRII~\eqref{RRII}. For all parameter values we have explored, if there exists a corresponding Kerr-AdS$_5$ black hole with the same volume and conserved charges, it has a larger area than the corresponding supergravity solution. Our results provide strong numerical support for the validity of~\eqref{RRII}.

In $D = 7$ gauged supergravity, there are exact rotating black hole solutions with three independent angular momenta and equal $U(1)$ charges~\cite{Chow:2007ts}. We study their extended thermodynamics for the first time in the Supplement. While studying the RRII for this black hole, computational constraints limited our verification to a few hundred random parameter sets. Despite this, no counter-examples emerged, even in solutions with pathologies like closed time-like curves.

We can make further analytical progress by noting a useful corollary of the strong RRII~\eqref{RRII} that follows when  the black hole of interest reduces to the Kerr-AdS black hole as some parameter $w \to 0$.\footnote{The RII and RRII do not require solutions to be connected to Schwarzschild-AdS or Kerr-AdS, as seen with our thin black rings.} If we further assume that the area is an analytic function of the parameter $w$, we can expand~\eqref{RRII} in the vicinity of the Kerr solution. The RRII implies in this limit that the first non-vanishing signed derivative of $A$ with respect to the parameter $w$ must be negative,
\be\label{perturbRRII} 
\lim_{w \to 0} {\rm sign}(w)^{n_\star} \left(\frac{\partial^{n_\star} A}{\partial w^{n_\star}} \right)_{M, J_i, V}  \le 0 \, ,
\ee
where $n_\star$ denotes the order of the first non-vanishing derivative. This inequality implies the RRII in some small neighbourhood of the Kerr solution where higher-order terms in a Taylor expansion can be neglected.
As such, this is a \textit{necessary condition} for the validity of conjecture~\eqref{RRII}, but it is not a sufficient condition. One advantage of~\eqref{perturbRRII} is that it can be applied directly to a particular black hole of interest and does not require a direct comparison with the Kerr solution.

For all black holes studied, we have proven that~\eqref{perturbRRII} holds. In every instance, the first derivative vanishes and the second is strictly negative. This reveals the Kerr-AdS black hole as a local entropy maximum, and confirms the RRII near the Kerr-AdS solution. 

\section{Bounds on the Isoperimetric Ratio}

It would be of interest to write~\eqref{RRII} as an explicit correction to the bound satisfied by the isoperimetric ratio $\mathcal{R}$. By making reference to the physical parameters of only a single black hole, such an inequality would apply in cases where a Kerr-AdS black hole may not exist for the specified values of $(M, J_i, V)$.

The first step toward constructing such an inequality would be to find a relationship amongst the physical parameters of the Kerr-AdS black hole of interest. Since the conjecture~\eqref{RRII} involves only $A$, $M$, $J_i$ and $V$ the necessary relationship would be a function of these parameters $f(A, M, J_i, V)$ such that $f(A, M, J_i, V) = 0$. In high dimensions obtaining such a relationship would generically require solving a polynomial of degree greater than four. This prevents us from presenting in the general case RRII that explicitly involves the isoperimetric ratio $\mathcal{R}$. However, below we will present weaker versions of~\eqref{RRII} that are similar to~\eqref{RII}. 

One exception is the four-dimensional case. There it is straight-forward to obtain a function of the relevant parameters that vanishes for Kerr-AdS,
\be\label{RRII_function} 
f(M, A, V, J) \equiv 36 \pi M^2 V^2 - M^2 A^3 - 64 \pi^3 J^4 \, ,
\ee
which we have implicitly made use of earlier in~\eqref{kerrads4}. 
In terms of this function, the inequality~\eqref{RRII} becomes the statement $f(A, M, V, J) \ge 0$. Some algebra allows this inequality to be expressed directly in terms of the isoperimetric ratio. The simplest way to do so yields,
\be\label{RRII4d} 
\mathcal{R} \ge \left[1 -\left( \frac{4 \pi J^2}{3M V} \right)^2\right]^{-1/6} \, .
\ee
The inequality is saturated for the Kerr-AdS black hole, and obviously reduces to~\eqref{RII} when $J \to 0$. In four dimensions, \eqref{RRII4d} is completely equivalent to~\eqref{RRII}.

In higher dimensions, it is not in general possible to do the same,  because obtaining the identity $A_{\rm Kerr} = A_{\rm Kerr}(M, J_i, V)$ requires solving a polynomial of high degree. For these cases we present reverse isoperimetric inequalities of intermediate strength, that is, strictly stronger than~\eqref{RII} but weaker than~\eqref{RRII}. The advantage of these intermediate inequalities is that they involve only quantities defined for one solution. Therefore, the intermediate inequalities have a potentially larger domain of applicability compared to the RRII. 

Due to differences in rotating black holes across even and odd dimensions, the intermediate inequalities also vary. We have the following in even and odd dimensions.

\begin{conjecture2-1*}[Intermediate RRII: even $D$]
\label{weakRRIIeven}
For a black hole of mass $M$, angular momenta $J_i$, area $A$ and thermodynamic volume $V$, the following inequality holds
\be 
{\cal R}^{D-1}
\ge\left[1-\left\{\frac{2\pi(D-2) J_{\min}^2}{(D-1)MV}\right\}^2\right]^{-1/2}
\ee
where $J_{\rm min} \equiv {\rm min} \{|J_i|\}$ and $\mathcal{R}$ is  defined in~\eqref{RII}. If it happens that there is only a single non-zero angular momentum, call it $J$, then we can further say
\ba\label{RRIIsinglespinEven}
{\cal R}^{D-1}
\ge  \left[1-\left\{\frac{8\pi}{(D-2)(D-1)}\frac{J^2}{MV}\right\}^2\right]^{-1/2}.
\ea
\end{conjecture2-1*}

\begin{conjecture2-2*}[Intermediate RRII: odd $D$]
\label{weakRRIIodd}
For a black hole of mass $M$, angular momenta $J_i$, area $A$ and thermodynamic volume $V$, the following inequality holds
\begin{align} 
\label{RRIIgeneralspinOdd}
{\cal R}^{D-1} \ \ge& \ \left(1- \frac{2\pi J_{\min}^2}{MV}\right)^{-(D-3)/[2(D-2)]} 
\nonumber\\
&\times\left(1+\frac{2\pi(D-3)}{(D-1)} \frac{J_{\min}^2}{MV}\right)^{-(D-1)/[2(D-2)]}
\end{align}
where $J_{\rm min} \equiv {\rm min} \{|J_i|\}$ and $\mathcal{R}$ is  defined in~\eqref{RII}. If it happens that there is only a single non-zero angular momentum, then we can further say
\begin{align}
\label{RRIIsinglespinOdd}
{\cal R}^{D-1}
 &\ge  \left[1- \frac{4\pi}{(D-1)(D-2)}\frac{J^2}{MV}\right]^{-(D-3)/[2(D-2)]}
\nonumber\\ 
&\times \left[1+\frac{4\pi (D-3)}{(D-2)(D-1)^2}\frac{J^2}{MV}\right]^{-(D-1)/[2(D-2)]}.
\end{align}
\end{conjecture2-2*}
Conjecture~2-1 is saturated for the equal-spinning Kerr-AdS black holes in even dimensions, while Conjecture~2-2 is saturated for the odd-dimensional Schwarzschild-AdS black holes. That they hold as inequalities for the general Kerr-AdS solutions is proven in the Supplement.

Numerically checking Conjectures~2-1 and 2-2 is more efficient than verifying the stronger~\eqref{RRII}. For the five-dimensional charged and rotating solution of minimal gauged supergravity, Conjecture~2-2 has been confirmed for approximately $10^{7}$ parameter sets. Similarly, in the seven-dimensional case with equal charges, we've confirmed it for approximately $10^{5}$ sets, providing strong evidence toward the validity of the conjecture.

Intriguingly, Conjectures~2-1 and 2-2 appear to hold beyond black holes with spherical horizon topology. We have checked these conjectures against the thin AdS black ring in all dimensions $D \ge 5$~\cite{Caldarelli:2008pz}. The thin black ring has horizon topology $\mathbb{S}^1 \times \mathbb{S}^{D-3}$ with the $\mathbb{S}^1$ characterized by the radius $R$ and the $\mathbb{S}^{D-3}$ characterized by the radius $r_0$. The ring is thin in the sense that $r_0 \ll {\rm min} \left\{R, \ell\right\}$, where $\ell$ is the AdS curvature radius. In particular, this means that the ratio $r_0/R \ll 1$ always. The thermodynamics of the thin black ring was explored in~\cite{Caldarelli:2008pz} and its extended thermodynamics in~\cite{Altamirano:2014tva}. The latter showed the RII's validity for the ring, with the isoperimetric ratio~\eqref{RII} greatly exceeding 1 due to $r_0/R \ll 1$ --- c.f.~ section~6 of that work.

Here we can prove analytically that the Conjectures~2-1 and~2-2 hold for the thin black ring. The solution has a single non-vanishing angular momentum, and so in each case it is the second inequality that applies. The key detail is the expression for the ratio
\be 
\frac{J^2}{M V} = \frac{(D-1) \left[ 1 + (D-2) \mathsf{R}^2 \right] \left[D-3 + (D-2) \mathsf{R}^2 \right]}{8 \pi (D-2) \left(1+ \mathsf{R}^2 \right)^2}
\ee
where we have introduced the notation $\mathsf{R} \equiv R/\ell$. This ratio is monotonically increasing, ranging from $J^2/(MV) = (D^2-4 D + 3)/[8 \pi (D-2)]$ for $\mathsf{R} = 0$ to $J^2/(MV) = (D^2-3 D + 2)/(8 \pi)$ in the limit $\mathsf{R} \to \infty$. This ratio is bounded and is order-one as a function of $R/r_0$. This fact ensures the intermediate RRII always holds (see Supplement).

\section{Conclusions}

The original reverse isoperimetric inequality appears to be part of a hierarchy, reminiscent of the nesting of Penrose inequalities for rotating and charged black holes. Within this hierarchy, the original RII is the least restrictive, and we have presented strong evidence for a more stringent version applicable to rotating black holes. This hints at an intricate link between thermodynamic volume and black hole entropy, revitalizing the initial conjecture and opening up fresh avenues of inquiry.

Here we have focused on the case of asymptotically AdS black holes in $D \ge 4$ dimensions, but similar questions could be explored in a variety of other contexts. For example, we anticipate an extension of this result to hold for de Sitter black holes and cosmological horizons,  along the lines of~\cite{Dolan:2013ft}. Furthermore, it is known that Misner strings possess thermodynamic volume~\cite{Bordo:2019tyh} and it may be possible to formulate a version of the (R)RII that applies to spacetimes with NUT charge. In  all cases it is natural to expect further possible extensions of the RII that incorporate angular momentum, charge, or possibly both charge and angular momentum. Our work has utilized examples of black holes with additional conserved charges. Analysing the validity of the new conjecture for hairy black holes, similar to~\cite{Astefanesei:2023sep}, would be an important step. Finally, understanding the holographic interpretation of both the RII and RRII would be worth further study. Using the framework of~\cite{Visser:2021eqk,Ahmed:2023snm} it should be possible to address this question in concrete terms. 

Another question concerns uniqueness. While both the Schwarzschild-AdS and Reissner-Nordstr{\"o}m-AdS black holes saturate the RII,  in $D = 4$ only the Kerr-AdS black hole saturates the RRII. This hints that saturation may occur \textit{only} for Kerr-AdS.

Ultimately, since the RRII is stronger than the original RII, finding a counter-example might be simpler. Such a counter-example would clarify the conditions for the (R)RII's validity and strengthen its mathematical foundation.


\begin{acknowledgments}

We are grateful to Roberto Emparan, David Kubiz{\v n}{\' a}k,  Hari Kunduri, Robert Mann, Jorge Rocha and Tetsuya Shiromizu for useful discussions, suggestions, and comments. MA is grateful to Keisuke Izumi, Shinji Mukohyama, Tetsuya Shiromizu, Takahiro Tanaka, Yoshimune Tomikawa, and Hirotaka Yoshino for their continuous encouragements and useful suggestions. MA would also like to thank Oversea Visiting Program for Young Researchers in ExU collaborations. RAH is grateful to the
Department of Applied Mathematics and Theoretical Physics at the University of Cambridge for hospitality during the completion of part of this work. The work of MA is supported by Grant-Aid for Scientific Research from Ministry of Education, Science, Sports and Culture of Japan (No. 21H05182 and 21H05189), the ANRI Fellowship, Grant-in-Aid for JSPS Fellows (No. 22J20147 and 22KJ1933), and JSPS Overseas Challenge Program for Young Researchers. AMF is supported by
ERC Advanced Grant GravBHs-692951, MICINN grant
PID2019-105614GB-C22, AGAUR grant 2017-SGR 754,
and State Research Agency of MICINN through the
“Unit of Excellence Mar\'{i}a de Maeztu 2020-2023”
award to the Institute of Cosmos Sciences (CEX2019-
000918-M). The work of RAH received the support of a fellowship from ``la Caixa” Foundation (ID 100010434) and from the European Union’s Horizon 2020 research and innovation programme under the Marie Skłodowska-Curie grant agreement No 847648” under fellowship code LCF/BQ/PI21/11830027.

\end{acknowledgments} 

\twocolumngrid

\bibliography{thebib}

\newpage 
\onecolumngrid
\begin{center}
    {\Large \bf Supplemental Material}
\end{center}

\maketitle 

\section{Extended Thermodynamics}

To clarify the definitions of the quantities in our inequalities, we will review the geometric foundation of extended thermodynamics. Thermodynamic volume has a geometric definition in terms of Komar integrals. Consider a black hole spacetime with stationary Killing field $k$ and co-rotating Killing field $\xi = k + \Omega_i m_i$ where $m_i$ are the rotational Killing fields. One then defines a regularized Komar mass, thermodynamic volume, and angular momenta as~\cite{Kastor:2009wy, Cvetic:2010jb}
\begin{align} 
M &= - \frac{(D-2)}{16 \pi  (D-3)} \int_{S_\infty} \left(\star {\rm d} k + \frac{4 \Lambda}{(D-2)} \star \omega_\xi \right) \, ,
\\ 
V &= -  \int_{\mathscr{H}} \star \omega_\xi \, , \quad J_i = \frac{1}{16 \pi } \int_{S_\infty} \star {\rm d} m_i \, .
\end{align}
Here $S_\infty$ corresponds to a surface where a bulk time-slice intersects infinity, $\mathscr{H}$ is a horizon cross-section, and $\omega_\xi$ is the Killing potential of the co-rotating Killing field defined by
\be 
\xi = \star {\rm d}  \star \omega_\xi \, .
\ee
A Killing potential can always be defined (at least locally) since any Killing vector automatically satisfies ${\rm d} \star \xi = 0$. The Killing potential regulates the Komar integral for the mass, making it finite in asymptotically AdS spacetimes. Note that the angular momenta do not require regularization.

As is obvious from its definition, the Killing potential is not unique as it has a gauge ambiguity. This gauge ambiguity has a clear physical interpretation: it corresponds to a choice of ground state $M = 0$. Once a ground state is chosen, the mass and thermodynamic volume are then determined in tandem. In practice, the gauge for the Killing potential is chosen so that the mass agrees with the Ashtekar-Magnon-Das (AMD) mass~\cite{Ashtekar:1984zz, Ashtekar:1999jx, Gibbons:2004ai}, which amounts to assigning the AdS vacuum zero mass and zero thermodynamic volume. The AMD mass agrees with that obtained by holographic renormalization up to Casimir energy terms~\cite{Ashtekar:1999jx, Hollands:2005wt}.\footnote{See the appendix of~\cite{Durgut:2022xzw} for an explicit example.} In the original RII and our generalizations of it that follow, it is always to be understood that the gauge of $\omega_\xi$ has been fixed so that the Komar mass agrees with the AMD mass.

\section{Metrics \& Thermodynamic Potentials}

In this appendix, we collect the technical details for the black hole solutions and their (extended) thermodynamic properties needed to perform the analysis in the main text. Note that here we use notation that matches the original references. Therefore, in the various subsections that follow the same notation may be used to refer to different objects. 

\subsection{Kerr-AdS Black Hole}

The metric for Kerr-AdS was first constructed in~\cite{Carter:1968ks} in four-dimensions. It was much later generalized to five dimensions~\cite{Hawking:1998kw}, and subsequently to all dimensions~\cite{Gibbons:2004uw}.  Its extended thermodynamics was studied in~\cite{Cvetic:2010jb}.  The metric is written as
\ba \label{KerrAdS_met}
ds^2&=&-\frac{W\rho^2}{l^2}d\tau ^2+\frac{2m}{U} \Bigl(W d\tau -\sum_{i=1}^{N} \frac{a_i \mu_i ^2 d\varphi _i}{\Xi _i}\Bigr)^2+\frac{U dr^2}{F-2m}\nonumber\\
&+&\sum_{i=1}^{N} \frac{r^2+a_i^2}{\Xi _i} \mu_i ^2 d\varphi _i^2+\sum_{i=1}^{N+\varepsilon}\frac{r^2+a_i ^2}{\Xi _i} d\mu _i ^2
-\frac{1}{W \rho^2}\Bigl(\sum_{i=1}^{N+\varepsilon}\frac{r^2+a_i ^2}{\Xi _i} \mu_i d\mu_i\Bigr)^2\,,
\ea
where
\begin{align}
\rho^2 &= r^2+l^2\,,\quad W=\sum_{i=1}^{N+\varepsilon}\frac{\mu _i^2}{\Xi _i}\,,\quad U= \sum_{i=1}^{N+\varepsilon} \frac{\mu _i^2 r^\varepsilon}{r^2+a_i^2} \prod _{j=1} ^N (r^2+a_j^2)\,,\quad
F = r^ {\varepsilon -2} \frac{\rho^2}{l^2}\prod_{i=1}^N (r^2+a_i^2)\,,\quad \Xi_i=1-\frac{a_i^2}{l^2}\,.\quad
\end{align}
$\varepsilon$ is 1 for even dimensions, and 0 for odd dimensions. $N$ is the number of independent rotations, which is given by $N=(D-1-\varepsilon)/2$.
Coordinates $\mu_i$ are not independent, and obey $\sum_i \mu_i^2=1$.
For even-dimensional case, $a_{N+1}$ is set to be zero.
The horizon radius $r_+$ is determined as the largest solution to $F-2m=0$. The relevant thermodynamic quantities here are the following:
\begin{align}
A &=\frac{\omega _{D-2}}{r_+^{1-\varepsilon}}\prod_{i=1}^N
\frac{a_i^2+r_+^2}{\Xi_i}\,,\quad
M = \frac{m \omega _{D-2}}{4\pi (\prod_{j} \Xi_j)}\Bigl(\sum_{i=1}^{N}{\frac{1}{\Xi_i}-\frac{1-\varepsilon }{2}}\Bigr)\,,\\
J_i &=  \frac{a_i m \omega _{D-2}}{4\pi \Xi_i (\prod_j \Xi_j)}\label{Kerr_MJ}\,,\quad V=\frac{r_+A}{D-1}+\frac{8\pi}{(D-1)(D-2)}\sum^N_{i=1} a_i J_i\,.
\end{align}

\subsection{Kerr-Newman-AdS Black Hole}

The Kerr-Newman-AdS black hole was first constructed in~\cite{Carter:1968ks}, and its  extended thermodynamics was studied in~\cite{Caldarelli:1999xj, Gunasekaran:2012dq}. Further details can be found in those references. The metric and gauge field take the form
\begin{align} 
ds^2 &= -\frac{\Delta_{\rm KN}}{\rho^2} \left[dt - \frac{a \sin^2\theta}{\Xi}d\phi \right]^2 + \frac{\rho^2}{\Delta_{\rm KN}} dr^2 + \frac{\rho^2}{S} d\theta^2 + \frac{S \sin^2\theta}{\rho^2}\left[adt - \frac{r^2+a^2}{\Xi} d\phi \right] \, ,
\\
\mathcal{A} &= - \frac{q r}{\rho^2} \left[dt - \frac{a \sin^2\theta}{\Xi} d\phi \right] \, ,
\end{align}
where
\begin{align}
\rho^2 &= r^2 + a^2 \cos^2\theta \, , \quad \Xi = 1 - \frac{a^2}{\ell^2} \, , \quad S = 1 - \frac{a^2}{\ell^2} \cos^2\theta \, , \quad \Delta_{\rm KN} = (r^2+a^2)\left(1+\frac{r^2}{\ell^2}\right) - 2 m r + q^2 \, .   
\end{align}
The event horizon is located at the largest root of $\Delta_{\rm KN}(r_+) = 0$. The relevant thermodynamic potentials are given by
\begin{align}
    A &= \frac{4\pi(r_+^2+a^2)}{\Xi} \, , 
    \quad 
    M = \frac{m}{\Xi^2} \, , 
    \quad 
    Q = \frac{q}{\Xi} \, , 
    \quad 
    J = \frac{am}{\Xi^2} \, , 
    \quad
    V = \frac{2 \pi}{3} \frac{(r_+^2+a^2) (2 r_+^2 \ell^2 + a^2 \ell^2 - r_+^2 a^2) + \ell^2 q^2 a^2}{\ell^2 \Xi^2 r_+} \, .
\end{align}
Here $A$ is the area of the horizon, $M$ is the mass, $J$ is the angular momentum, $Q$ is the electric charge, and $V$ is the thermodynamic volume. Note that requiring a non-singular metric means that $\Xi > 0$ which limits $|a| < \ell$.

\subsection{Charged and Rotating AdS C-metric}

The extended thermodynamics of the charged and rotating AdS C-metric was studied in~\cite{Anabalon:2018qfv}, and we have also made use of the results of~\cite{Gregory:2019dtq}. We refer to those references for additonal details and derivations. The metric and gauge field are given by
\begin{align}
    ds^2 &= \frac{1}{H^2} \left\{-\frac{f}{\Sigma} \left[\frac{dt}{\alpha} - a \sin^2 \theta \frac{d\phi}{K} \right]^2 + \frac{\Sigma}{f} dr^2 + \frac{\Sigma r^2}{h} d\theta^2 + \frac{h \sin^2\theta}{\Sigma r^2} \left[\frac{a dt}{\alpha} - (r^2+a^2) \frac{d\phi}{K}\right]^2 \right\} \, , 
    \\
    \mathcal{A} &= - \frac{q}{\Sigma r} \left[ \frac{dt}{\alpha} - a \sin^2\theta \frac{d\phi}{K} \right] \, ,
\end{align}
where
\begin{align}
    f &= (1-A^2 r^2) \left[1- \frac{2m}{r} + \frac{a^2+q^2}{r^2} + \frac{r^2+a^2}{\ell^2} \right] \, , 
    \quad 
    h = 1 + 2 m A \cos \theta + \left[A^2(a^2+q^2) - \frac{a^2}{\ell^2} \right] \cos^2\theta \, ,
    \\
    \Sigma &= 1 + \frac{a^2}{r^2} \cos^2 \theta \, , 
    \quad 
    H = 1 + A r \cos \theta \, .
\end{align}
The parameter $A$ is related to the acceleration of the spacetime. The event horizon is located at the largest root of $f(r_+) = 0$. In the limit of slow acceleration $A \ell < 1$, which is the focus here, there is no acceleration horizon. 

The thermodynamic potentials in terms of the parameters of the metric appear in Eq.~(11) of~\cite{Anabalon:2018qfv}. In our work, we have found the relationships (10) and (11) of~\cite{Gregory:2019dtq} more convenient. The relevant expressions are the following:
\begin{align}
M^2 &= \frac{\Delta_{\rm C} S}{4 \pi} \left[ \left(1 + \frac{\pi Q^2}{\Delta_{\rm C} S} + \frac{8 P S}{3 \Delta_{\rm C}} \right)^2  + \left(1 + \frac{8 P S}{3 \Delta_{\rm C}} \right)\left(\frac{4 \pi^2 J^2}{\Delta_{\rm C}^2 S^2} - \frac{3 C^2 \Delta_{\rm C}}{2 P S} \right)\right] \, ,
\\ 
V &= \frac{2 S^2}{3 \pi M} \left[\left(1 + \frac{\pi Q^2}{\Delta_{\rm C} S} + \frac{8 P S}{3 \Delta_{\rm C}} \right) + \frac{2 \pi^2 J^2}{\Delta_{\rm C}^2 S^2} + \frac{9 C^2 \Delta_{\rm C}^2}{32 P^2 S^2} \right] \, .
\end{align}
Here $M$ is the mass, $V$ is the thermodynamic volume, $Q$ is the electric charge, $J$ is the angular momentum and $S = A/4$ is the horizon entropy. The pressure is $P = -\Lambda/(8\pi)$. The remaining two parameters $C$ and $\Delta_{\rm C}$ are related to the average and differential conical deficits of the spacetime, respectively --- see (9) of~\cite{Gregory:2019dtq}. Specifically, in our work in the main text, it was important to note that the parameter $C \ge 0$ and $\Delta_{\rm C} \ge 0$. In the main text, we also used the shorthand 
\be 
x \equiv \frac{8 P S}{\Delta_{\rm C}} \, ,
\ee
which is introduced in (13) of~\cite{Gregory:2019dtq} and used in (17) of~\cite{Gregory:2019dtq}.

\subsection{Pairwise-Equal Charge $D=4$ Gauged Supergravity Black Hole}

These black holes were constructed in~\cite{Chong:2004na} and we refer the reader to that manuscript for the explicit form of the metric. The extended thermodynamics was studied in~\cite{Cvetic:2005zi, Cvetic:2010jb} and here we present the relevant thermodynamic quantities:
\begin{align}
    M &= \frac{m+q_1+q_2}{\Xi^2} \, , 
    \quad 
    A = \frac{4 \pi (r_1 r_2 + a^2)}{\Xi} \, , 
    \quad
    J = \frac{a (m+q_1+ q_2)}{\Xi^2} \, , 
    \quad 
    Q_1 = Q_2 = \frac{\sqrt{q_1(q_1+m)}}{2 \Xi} \, , 
    \quad 
    \\ 
    Q_3 &= Q_4 = \frac{\sqrt{q_2(q_2+m)}}{2 \Xi}  \, ,
    \\
    V &= \frac{2 \pi \left[2 r_+ \Xi (r_+ + q_1 + q_2)(a^2+r_1 r_2) +  a^2 (2r_+(q_1 + q_2) + r_+^2 + g^2r_1^2 r_2^2 + a^2 (1+g^2 r_1 r_2)\right]}{3 \Xi^2 r_+} 
\end{align}
where 
\be 
r_1 = r_+ + 2 q_1 \, , \quad r_2 = r_+ + 2 q_2 \, , \quad \Xi = 1 - a^2 g^2 \, , \quad g \equiv \frac{1}{\ell} \, .
\ee
Here, $\ell$ is the cosmological length scale and the horizon radius $r_+$ is determined as the largest root to the following,
\be 
\Delta_4 = r^2 + a^2 - 2 m r + g^2 r_1 r_2(r_2 r_2 +a^2 ) \, .
\ee
Sensible solutions require real values for the charges $Q_1$ and $Q_3$, while also demanding $\Xi > 0$, or in other words, $|a| < \ell$.

\subsection{Charged and Rotating Black Holes of $D = 5$ Minimal Gauged Supergravity} \label{sub:D5_sugra}

This metric was first constructed in~\cite{Chong:2005hr} and its extended thermodynamics was studied in~\cite{Cvetic:2010jb}. In this case, the metric is sufficiently complicated that we do not present it here and instead refer to the original references. The necessary thermodynamic data required here is the following:
\begin{align}
    M &= \frac{m \pi \left( 2 \Xi_a + 2 \Xi_b - \Xi_a\Xi_b\right) + 2 \pi q a b g^2 \left(\Xi_a + \Xi_b \right)}{4 \Xi_a^2 \Xi_b^2} \, , 
    \quad 
    A = \frac{2 \pi^2 \left[(r_+^2 + a^2)(r_+^2+b^2) + a b q \right]}{\Xi_a \Xi_b r_+} \, ,
    \\ 
    J_a &= \frac{\pi \left[2 a m + q b (1+a^2 g^2) \right]}{4 \Xi_a^2 \Xi_b} \, ,
    \quad 
    J_b = \frac{\pi \left[2 b m + q a (1+b^2 g^2) \right]}{4 \Xi_a \Xi_b^2} \, , 
    \quad 
    Q = \frac{\sqrt{3} \pi q}{4 \Xi_a \Xi_b} \, ,
    \\ 
    V &= \frac{\pi^2 (r_+^2+a^2)(r_+^2+b^2)\left[3 r_+^2 + (a^2+b^2)(1-2 g^2 r_+^2)+ a^2 b^2 g^2(g^2r_+^2- 2) \right]}{6 \Xi_a^2 \Xi_b^2 r_+^2}
     \nonumber\\
     &- \frac{\pi^2 a b q \left[-2 r_+^2 + (a^2+b^2)(g^2r_+^2-1) + 2 a^2 b^2 g^2 \right] }{3 \Xi_a^2 \Xi_b^2 r_+^2} + \frac{\pi^2 q^2 \left[b^2+a^2 - 2 a^2 b^2 g^2 \right]}{6 \Xi_a^2 \Xi_b^2 r_+^2}
\end{align}
where we have
\be 
\Xi_a = 1 - a^2 g^2 \, , \quad \Xi_b = 1 - b^2 g^2 \, , \quad g \equiv \frac{1}{\ell}
\ee
with $\ell$ the cosmological length scale. 
The horizon radius is determined as the largest root of $\Delta_5(r_+)$ where 
\be 
\Delta_5 = \frac{(r^2+a^2)(r^2+b^2)(1+g^2r^2) + q^2 + 2 a b q}{r^2} - 2 m \, .
\ee
An important consideration in our analysis was ensuring that the value of $r_+$ generated randomly is in fact the largest root of $\Delta_5$. Caution is required in the following sense. If we solve the above condition $\Delta_5(r_+) = 0$ for $m$ and then substitute this value back into $\Delta_5$, we obtain
\be 
\Delta_5 = \frac{(r-r_+)(r+r_+) \left[(a b + q - r r_+)(a b + q + r r_+) - g^2 r^2 r_+^2 (a^2+b^2+r^2+r_+^2)\right]}{r^2 r_+^2} \, .
\ee
Obviously $\Delta_5 = 0$ for $r = r_+$. However, the term in square brackets contains a mixture of positive and negative terms. Therefore, for some parameter values this term in square brackets could have zeros for some value of $r$. If such a zero occurs, it could happen that this zero occurs at a larger value of $r$ than $r_+$. In such a case, $r = r_+$ actually represents the inner horizon, rather than the event horizon. Taking care of this technical point is important in verifying the RRII.

\subsection{Rotating and Equal Charge Black Holes in $D=7$ Gauged Supergravity}

This metric is discussed in~\cite{Chow:2007ts}. The solution includes three independent angular momenta depending on the rotational parameters $(a_1,\ a_2,\ a_3)$, equal charges and an arbitrary gauge coupling constant $g$. As in Subsection~(\ref{sub:D5_sugra}), we report here the relevant thermodynamic quantities: 
\begin{align}
    M &= \frac{\pi^2}{8 \Xi_1 \Xi_2 \Xi_3} \left[ \sum_{i} \frac{2 m}{\Xi_{i}} - m + \frac{5 q}{2} + \frac{q}{2} \sum_{i} \left( \sum_{j \neq i} \frac{2 \Xi_{j}}{\Xi_{i}} - \Xi_{i} - \frac{2 (1+2 a_1 a_2 a_3 g^3)}{\Xi_{i}}\right)\right]\,,\\
    A &= \frac{\pi^3}{\Xi_1 \Xi_2 \Xi_3 r_{+}} \left[ (r_{+}^2 + a_1^2) (r_{+}^2 + a_{2}^2)(r_{+}^2 + a_{3}^2) + q (r^2_{+} - a_{1} a_{2} a_{3} g)\right]\,,\\
    J_{i} &= \frac{\pi^2 m}{4 \Xi_{1} \Xi_{2} \Xi_{3} \Xi_{i}} \left[ a_{i} c^2 - s^2 g \left( \Pi_{j \neq i} a_{j} + a_{i} \sum_{j \neq i} a_{j}^2 g + a_{1} a_{2} a_{3} a_{i} g^2\right) \right], \,\, Q = \frac{\pi^2 m s c}{\Xi_{1} \Xi_{2} \Xi_{3}}
\end{align}
with 
\begin{align}
    \Xi_{i} = 1 - a_{i}^2 g^2, \quad c^2 = 1+\frac{q}{2 m},\quad s^2 = \frac{q}{2m},\quad g\equiv \frac{1}{\ell}.
\end{align}

The extended thermodynamics of this solution has not been previously examined, and we present here the thermodynamic volume (calculated by using the Smarr relation and the first law~\cite{Cvetic:2010jb})
\begin{small}
\begin{align}
    V &= \frac{\pi ^3}{\prod _{i=1}^3 \Xi _i^2} \left\{
    \frac{16 q r^2 \left(q r^2-A\right) \left(\prod _{i=1}^3 \Xi _i\right)}{30 g^2 r^2 \left[q \left(r^2-g \prod _{i=1}^3 a_i\right)+\prod _{i=1}^3 \Tilde{B}_i\right]} - 
    \frac{\left(\prod _{i=1}^3 \Xi _i\right)
    }{6 g^2 r^2} \left[  g^2 q^2 -r^2 \left(g^2 r^2+1\right) \left(2 r^2 \sum _{i=1}^3 a_i^2+\Tilde{F}+3 r^4\right) + \right.
    \right. \nonumber\\
    &  \left.-2 g q \left(\prod _{i=1}^3 a_i+g r^4\right)+\prod _{i=1}^3 \Tilde{B}_i\right]- \frac{q \prod _{i=1}^3 \Xi _i}{3 g^2} + \frac{1}{15 g^2 r^2}\left[\left(
    4 g^2 \sum _{i=1}^3 a_i^2 -\frac{g^4}{2} \left( \sum _{i=1}^3 \sum _{j=1}^3 a_i^2 a_j^2+5 \sum _{i=1}^3 a_i^4\right) +4 g^3\left(3  + \Tilde{F} g^4 \right) \prod _{i=1}^3 a_i
    \right. \right.\nonumber
    \\
    &   \left.\left.
      + \sum _{i=1}^3 a_i^2 \left( g^8 \prod _{i=1}^3 a_i^2  -8 g^5 \prod _{i=1}^3 a_i  \right)
      +g^6 +\left(\sum _{i=1}^3 \sum _{j \neq i}^3 a_i^4 a_j^2-6 \prod _{i=1}^3 a_i^2\right) - 3
     \right) q r^2 \right. \left. - \Tilde{A} \left(\prod _{i=1}^3 \Xi _i -2 \left(3-2 g^2 \sum _{i=1}^3 a_i^2+F g^4\right)\right)\right]+\nonumber\\
    & +\frac{ \sum_{i=1}^{3}  \prod_{j \neq i} \Xi_{j} \left\{ q g \prod_{j \neq i} a_{j} - a_{i} \left[
    g^{2}q r^2 + \prod_{j \neq i} \Tilde{B}_{j}\left(1 + g^2 r^2 \right)
    \right] \right\}   
    }{6 g^2 r^2 \left[q \left(r^2-g \prod _{i=1}^3 a_i\right)+\prod _{i=1}^3 \Tilde{B}_i\right]}
    \left\{
    g \prod_{\substack{i \neq j\\ k \neq i,j}} (a_{j} + a_{i} a_{k} g) q r^2 + a_{i} \left[
    2 q g \prod_{\ell=1}^3 a_{\ell} -g^2 q^2  +
    \right.
    \right.\nonumber\\
    &
    \left. \left.
    -g^2 q r^2 \left(\sum_{\ell =1}^3 a_{\ell}^2 + 2 r^2 \right) - \left(\prod_{\ell = 1}^3 \Tilde{B}_{\ell} \right) \left( 1 + g^2 r^2 \right)- q r^2
    \right]
    \right\},
\end{align}
\end{small}
with
\begin{align}
    \Tilde{A}&\equiv -g^2 q r^2 \left(\sum _{i=1}^3 a_i^2+2 r^2\right)+2 g q \left(\prod _{i=1}^3 a_i\right)-\left(g^2 r^2+1\right) \left(\prod _{i=1}^3 \Tilde{B}_i\right)-g^2 q^2, \\
    \Tilde{F} &\equiv \frac{1}{2} \left(\sum _{i=1}^3 \sum _{j=1}^3 a_i^2 a_j^2-\sum _{i=1}^3 a_i^4\right),\\
    \Tilde{B}_i &\equiv a_i^2+r^2.
\end{align}

As explained in~\cite{Chow:2007ts}, in the limit of $q \rightarrow 0$, the thermodynamic quantities coincide with the Kerr-AdS black holes in $D=7$~\cite{Gibbons:2004ai}. Note that this is also valid for the thermodynamic volume.

The same considerations made in Subsection~(\ref{sub:D5_sugra}), need to be done here where the horizon radius is given by the largest root of
\begin{align}
    \Delta_7 &= \frac{(1+g^2 r^2) \left[ (r^2+a_{1}^2)(r^2+a_{2}^2)(r^2+a_{3}^2) + q g^2 (2 r^2 + a_{1}^2 + a_{2}^2 + a_{3}^2) \right] - 2 q g a_{1} a_{2} a_{3} + q^2 g^2 }{r^2} - 2 m.
\end{align}

\subsection{AdS Thin Black Ring}

The AdS thin black ring in $D\ge5$ was first constructed in~\cite{Caldarelli:2008pz}, and its extended thermodynamics was analysed in~\cite{Altamirano:2014tva}. The main idea of the construction is to bend a boosted thin black string of thickness $r_0$ into a circle of radius $R$. In the asymptotic region, the thin black ring is then regarded as a distributional source in a global AdS background,
\be 
ds^2 = \left(1+\frac{\rho^2}{\ell^2} \right) dt^2 + \left(1+\frac{\rho^2}{\ell^2} \right)^{-1} d\rho^2 + \rho^2 \left[d\Theta^2 +  \sin^2 \Theta d\Omega_{D-4} + \cos^2\Theta d\psi^2\right] \, ,
\ee
placed at $\rho = R$ in the $\Theta = 0$ plane. The ring carries angular momentum along $\partial_\psi$ (i.e. in a single plane) which serves to balance the ring tension and gravitational potential. 

The thermodynamic quantities of the ring can be found in~\cite{Caldarelli:2008pz, Altamirano:2014tva}, and here we include only the necessary quantities to verify the results in the main text.
\begin{align}
    M &= \frac{(D-2)  \Omega_{D-3} r_0^{D-4} \ell}{8} \mathsf{R} \left(1+\mathsf{R}^2 \right)^{3/2} \, ,\quad
    J  = \frac{\Omega_{D-3} r_0^{D-4} \ell^2 \mathsf{R}^2}{8} \sqrt{\left(1 + (D-2) \mathsf{R}^2 \right) \left( D- 3 + (D-2)\mathsf{R}^2\right)} \, ,\label{BR_MJ}
     \\ 
    A &= 2\pi \Omega_{D-3} \ell r_0^{D-3}\mathsf{R} \sqrt{\frac{D-3 + (D-2)\mathsf{R}^2}{D-4}}
         \, ,\quad
    V = \frac{\pi \Omega_{D-3} r_0^{D-4}}{D-1} \mathsf{R}^3 \sqrt{1 + \mathsf{R}^2}  \, .\label{BR_AV}
\end{align}
Here, as in the main text, we have used the notation $\mathsf{R} \equiv R/\ell$, where $R$ is the radius of the $\mathbb{S}^1$ of the ring. The validity of the thin ring approximation requires that $r_0 \ll {\rm min} \left\{R, \ell \right\}$. No hierarchy of scales is assumed between $R$ and $\ell$, allowing for one to consider all ratios of $R/\ell$.

We now give a more detailed proof of the validity of the intermediate RRII inequalities in even and odd dimensions. First, as proven in~\cite{Altamirano:2014tva}
[by (6.15) (for $R\leq l$) and (6.16) (for $R> l$)], we have
\ba
{\cal R}^{D-1} \sim \left[\frac{\ell \mathsf{R}}{r_0} \right]^{(2D-5)/(D-2)} \left[\frac{1+\mathsf{R}^2}{D-3 + (D-2) \mathsf{R}^2}\right]^{1/[2(D-2)]} \label{R>>1}
&\gg&1,
\ea
where we ignore overall (positive) constants in the first line.   $\mathcal{R} \gg 1$  holds because the right-hand side always depends on positive powers of the ratio $R/r_0$, which is a large number within the thin ring approximation.

Due to Eqs.~\eqref{BR_MJ} and \eqref{BR_AV}, we have
\ba
\frac{J^2}{MV}&=&\frac{(D-1)\left[1+(D-2)\mathsf{R}^2\right]\left[D-3+(D-2)\mathsf{R}^2\right]}{8\pi(D-2)(1+\mathsf{R}^2)^2}.\label{J^2/MV} \label{JJ/MV}
\ea
Let us discuss odd-dimensional case and even-dimensional case separately.

\subsubsection{Even Dimensions}

In this section, we will check the following conjecture for the single spinning case:
\ba
{\cal R}^{D-1}
&\ge&\left[1-\left\{\frac{8\pi}{(D-2)(D-1)}\frac{J^2}{MV}\right\}^2\right]^{-1/2},\label{refiened_RII_1_evensin}
\ea
Let us evaluate the right-hand side of \eqref{refiened_RII_1_evensin}. For $\mathsf{R}\le 1$, ~\eqref{JJ/MV} gives us
\ba
\left[1-\left\{\frac{8\pi}{(D-2)(D-1)}\frac{J^2}{MV}\right\}^2\right]^{-1/2}&\le&\left[1-\left\{\frac{\left(D-1\right)\left(2D-5\right)}{4(D-2)^2}\right\}^2\right]^{-1/2}.\label{RHS_even1}
\ea
Note that 
the right-hand side of \eqref{RHS_even1} is real.
For $\mathsf{R}> 1$, we have 
\ba
\left[1-\left\{\frac{8\pi}{(D-2)(D-1)}\frac{J^2}{MV}\right\}^2\right]^{-1/2}&<&2\mathsf{R}.\label{RHS_even2}
\ea
These are always $\mathcal{O}(1)$ quantities in terms of the ratio $R/r_0$. Due to \eqref{R>>1}, \eqref{RHS_even1}, and \eqref{RHS_even2}, the conjecture of \eqref{refiened_RII_1_evensin} holds.

\subsubsection{Odd Dimensions}
For odd-dimensions, Conjecture~2-2 for an angular momentum in a single plane, reduces to
\ba
{\cal R}^{D-1}
&\ge& \left[1- \frac{4\pi}{(D-1)(D-2)}\frac{J^2}{MV}\right]^{-(D-3)/[2(D-2)]}\left[1+\frac{4\pi (D-3)}{(D-2)(D-1)^2}\frac{J^2}{MV}\right]^{-(D-1)/[2(D-2)]}.\label{refiened_RII_1_oddsin}
\ea
For the right-hand side of \eqref{refiened_RII_1_oddsin}, we obtain
\ba
&&\hspace{-20mm}\left[1- \frac{4\pi}{(D-1)(D-2)}\frac{J^2}{MV}\right]^{-(D-3)/[2(D-2)]}\left[1+\frac{4\pi (D-3)}{(D-2)(D-1)^2}\frac{J^2}{MV}\right]^{-(D-1)/[2(D-2)]}\nonumber\\
&=&\left[1- \frac{\left[1+(D-2)\mathsf{R}^2\right]\left[D-3+(D-2)\mathsf{R}^2\right]}{2(D-2)^2(1+\mathsf{R}^2)^2}\right]^{-(D-3)/[2(D-2)]}\nonumber\\
&&\hspace{35mm}\times\left[1+\frac{(D-3)\left[1+(D-2)\mathsf{R}^2\right]\left[D-3+(D-2)\mathsf{R}^2\right]}{2(D-1)(D-2)^2(1+\mathsf{R}^2)^2}\right]^{-(D-1)/[2(D-2)]}\nonumber\\
&<&2^{(D-3)/[2(D-2)]}\left[1+\frac{(D-3)}{2(D-1)}\right]^{-(D-1)/[2(D-2)]}.\label{RHS_odd}
\ea
This is always an $\mathcal{O}(1)$ quantity in terms of the ratio $R/r_0$. Due to \eqref{R>>1} and \eqref{RHS_odd}, we see that the conjecture of \eqref{refiened_RII_1_oddsin} holds.

\section{Technical Details for Higher-Dimensional `Intermediate' Inequalities}
\label{weakInequals}
Let us calculate the isoperimetric ratio ${\cal R}$ following the result in~\cite{Cvetic:2010jb}.
We introduce $z$ as  
\be
z \ \equiv \ \frac{1+r_+^2/l^2}{r_+^2}\sum_i \frac{a_i^2}{\Xi_i}\,,\label{def:z}
\ee
which can be rewritten as
\ba
z^{-1}&=&\left(\frac{8\pi}{D-1}\sum_{i} a_i J_i\right)^{-1} V-\frac{1}{D-2}.\label{z^-1}
\ea

Let us start from the even-dimensional case.
The isoperimetric ratio ${\cal R}$ is obtained in~\cite{Cvetic:2010jb} as 
\ba
{\cal R}^{D-1}
&=&\Bigl(1+\frac{z}{D-2}\Bigr)\Bigl(1+\frac{2z}{D-2}\Bigr)^{-1/2}\,,\label{RII_ratio_even}
\ea
which satisfies $d {\cal R}/dz\ge0$.
Due to \eqref{z^-1}, we obtain
\ba
z^{-1}&\le&\left(4\pi\frac{D-2}{D-1}a_{\min}  J_{\min} \right)^{-1} V-\frac{1}{D-2}\,,\label{z^-1_evengen}
\ea 
where $a_{\min}$ and $J_{\min}$ are the minima of $|a_i|$ and $|J_i|$, respectively. The equality in \eqref{z^-1_evengen} holds for the equally rotating case.
Due to \eqref{Kerr_MJ}, we have
\ba
M \ = \ \sum^N_{i=1}\frac{J_i}{a_i} \ \ge \ \frac{(D-2)J_{\min} }{2a_{\min} }\,,\label{J/a_evengen}
\ea
where, again, the equality holds for the equality rotating case.
Then, from \eqref{z^-1_evengen} and \eqref{J/a_evengen}, we obtain 
\ba
z^{-1}&\le&\left[2\pi\frac{(D-2)^2}{D-1}\frac{J_{\min}^2}{M} \right]^{-1} V-\frac{1}{D-2}.\label{ineq:z^-1_evengen}
\ea
By using \eqref{RII_ratio_even}, \eqref{ineq:z^-1_evengen} and $d {\cal R}/dz\ge0$, we have
\ba \label{RRii_even}
{\cal R}^{D-1}
&\ge&\left[1-\left\{\frac{2\pi(D-2) J_{\min}^2}{(D-1)MV}\right\}^2\right]^{-1/2}.\label{refiened_RII_1_evengen}
\ea
Similarly, for even-dimensional single spinning case, \eqref{z^-1_evengen} and \eqref{J/a_evengen} can be rewritten as 
\ba
z^{-1} \ = \ \left(\frac{8\pi}{D-1}a J \right)^{-1} V-\frac{1}{D-2}\,,\qquad
M \ = \ \frac{J}{a}.
\ea
After similar steps as for the general spinning case, we have
\ba\label{RRii_even_singly}
{\cal R}^{D-1}
 \ = \ \left[1-\left\{\frac{8\pi}{(D-2)(D-1)}\frac{J^2}{MV}\right\}^2\right]^{-1/2}.
\ea
The resemblance between~\eqref{RRii_even} and~\eqref{RRii_even_singly} seems to suggest that~\eqref{RRii_even} might be improved by replacing $J_{\min}$ by the average of the absolute value of all the black hole's angular momenta, i.e.,  $\Sigma_{i} | J_{i}|/(D/2 - 1 ) $. In this way, the denominator would be the number of distinct angular momenta that can exist in even $D$ dimension, and it is precisely the factor that, when squared, yields the coefficient that appears in~\eqref{RRii_even_singly}. However, this gives a weaker version of the conjecture being more restrictive than the one given by~\eqref{RRii_even}. Moreover, as we will see, it seems not to be possible to obtain the similar relation in odd dimensions.

In odd dimensions, on the other hand, the isoperimetric ratio ${\cal R}$ is calculated as \cite{Cvetic:2010jb}
\ba
{\cal R}^{D-1}
\ = \ \Big(1+ \frac{z}{D-2}\Big) \Big(1+ \frac{2z}{D-1}
\Big)^{-(D-1)/[2(D-2)]}\,,\label{RII_ratio_odd}
\ea
which gives us $d {\cal R}/dz \ge  0$.
Due to \eqref{Kerr_MJ}, we have
\ba 
\sum^N_{i=1}\frac{J_i}{a_i} 
\ = \ M+\frac{m\Omega_{D-2}}{8\pi\prod_{i}\Xi_i},\qquad
M \ \ge \ (D-2)\frac{m\Omega_{D-2}}{8\pi\prod_{j}\Xi_j}\,,\label{J/a_oddgen}
\ea
where the equality holds for non-rotating (Schwarzschild-AdS) case.
Then, from \eqref{Kerr_MJ} and \eqref{J/a_oddgen}, we obtain 
\ba
\frac{D-1}{D-2}M \ \ge \ \sum^N_{i=1}\frac{J_i}{a_i}  \ \ge \ \frac{D-1}{2}\frac{J_{\min}}{a_{\min}}.\label{inequ:J/a_oddgen}
\ea
Then, by \eqref{z^-1} and \eqref{inequ:J/a_oddgen}, we obtain
\ba
z^{-1}&\le&\left[2\pi(D-2) \frac{J_{\min}^2}{M}\right]^{-1} V-\frac{1}{D-2}.\label{ineq:z^-1_oddgen}
\ea
By $d{\cal R}/dz\ge0$, \eqref{RII_ratio_odd} and \eqref{ineq:z^-1_oddgen},  we have
\ba
{\cal R}^{D-1} \ \ge \ \left(1-2\pi \frac{J_{\min}^2}{MV}\right)^{-(D-3)/[2(D-2)]}\left(1+2\pi\frac{D-3}{D-1} \frac{J_{\min}^2}{MV}\right)^{-(D-1)/[2(D-2)]}.
\label{refiened_RII_1_oddgen}
\ea
Similarly, for odd-dimensional single spinning case, we obtain
\ba
{\cal R}^{D-1}
\ \ge \ \left[1- \frac{4\pi}{(D-1)(D-2)}\frac{J^2}{MV}\right]^{-(D-3)/[2(D-2)]}\left[1+\frac{4\pi (D-3)}{(D-2)(D-1)^2}\frac{J^2}{MV}\right]^{-(D-1)/[2(D-2)]},
\ea
where the equality holds for $J=0$.

\end{document}